\documentclass[12pt]{iopart}
\usepackage[utf8]{inputenc} 
\usepackage[T1]{fontenc}    
\usepackage{hyperref}       
\usepackage{url}            
\usepackage{booktabs}       
\usepackage{amsfonts}       
\usepackage{amsbsy}
\usepackage{nicefrac}       
\usepackage{microtype}      
\usepackage{xcolor}         

\usepackage{caption}
\usepackage{subcaption}

\usepackage{iopams}  
\usepackage{amsmath}

\usepackage{comment}
\usepackage{listings}
\usepackage{graphicx}
\usepackage{algorithm}
\usepackage{algorithmic}
\usepackage{amssymb}
\usepackage{listings}
\usepackage{bm, bbm}
\usepackage{bbold}
\usepackage{graphics}
\usepackage{tikz}
\usepackage{graphicx}

\newcommand{\iqnone}{\texttt{IQNx1}}
\newcommand{\iqnfour}{\texttt{IQNx4}}

\begin{document}

\title{Implicit Quantile Networks for Emulation in Jet Physics}

\author{Braden Kronheim$^1,$\footnote{corresponding author}, Ali Al Kadhim$^2$, Michelle P. Kuchera$^{3,4}$, Harrison B. Prosper$^2$, and Raghuram Ramanujan$^4$}

\address{$^1$ Dept.~of Physics, University of Maryland, College Park, MD 20742}
\address{$^2$ Dept.~of Physics, Florida State University, Tallahassee, FL 32306}
\address{$^3$ Dept.~of Physics, Davidson College, Davidson, NC 28035}
\address{$^4$ Dept.~of Mathematics and Computer Science, Davidson College, Davidson, NC 28035}
\ead{bkronhei@umd.edu}
\vspace{10pt}
\begin{indented}
\item[]\today
\end{indented}


\begin{abstract}

  The ability to model and sample from conditional densities is important in many physics applications. Implicit quantile networks (IQN) have been successfully applied to this task in domains outside physics.
  In this work, we illustrate the potential of IQNs as components of emulators using the simulation of jets as an example. Specifically, we use an IQN to map jets described by their 4-momenta at the generation level to jets at the event reconstruction level. The conditional densities emulated by our model closely match those generated by \texttt{Delphes}, while also enabling faster jet simulation. 
\end{abstract}
\vspace{2pc}
\noindent{\it Keywords}: Generative AI, Jet Physics, Large Hadron Collider

\submitto{Machine Learning: Science and Technology}
\maketitle
%

%
%
%

\section{Introduction}
High-fidelity simulators are of critical importance in many fields of science as they provide the connection between theoretical models and potential (and actual) observations. In high-energy physics, the simulation pipeline  comprises an event generator 
which encodes the theoretical predictions of particle interactions,
a detector simulator,
and event reconstruction which transforms low-level data to objects that model final state particles. 

Viewed abstractly, a high-energy physics simulator is a procedure that maps an event $\bm{x}$ comprising one set of particles to another event $\bm{y}$ comprising a different set of particles. 
A collection of such multi-level events is a point cloud approximation of expressions of the form

\begin{equation}
o(\bm{y}) = \int p(\bm{y} | \bm{x}) \, u(\bm{x}) \, d\bm{x},
\label{eq:map}
\end{equation}
each of which maps a  density $u(\bm{x})$, typically unobserved, to an observed density $o(\bm{y})$ via a  conditional density $p(\bm{y} | \bm{x})$, which, 
in general, is multi-dimensional. In some contexts, $p(\bm{y} | \bm{x})$ is called a \emph{response function}. The mapping $u(\bm{x}) \rightarrow o(\bm{y})$ is referred to as \emph{folding}, while the inverse mapping $o(\bm{y}) \rightarrow u(\bm{x})$ is referred to as \emph{unfolding}\footnote{ Unlike folding, unfolding is an ill-posed problem due to the information loss from folding. In order to render an unfolding procedure well-posed, information must be injected into the procedure via some form of regularization.}. 

An example of Eq.\,(\ref{eq:map}) is the mapping of a jet,
a collimated collection of particles (see, for example, \cite{osti_1263408}), prior to its interaction with a particle detector to the jet observed in the detector. Traditionally, the interaction of a jet of particles with a detector is modeled using a Monte Carlo method based on the widely used \texttt{GEANT4} toolkit~\cite{Ivanchenko:2003xp}. \texttt{GEANT4} provides  high-fidelity simulations of particle interactions with matter, but comes with a high computational cost. For this reason, experimental collaborations have devoted considerable effort to building  fast simulators (see, for example, \cite{Abdullin2011-tg, Aad2010-om}) in which the slower parts of the full (\texttt{GEANT4}-based) simulator are replaced by hand-coded parameterized approximations to the conditional densities $p(\bm{y} | \bm{x})$.

Unfortunately, hand coding of these densities for fast simulators is an error-prone and labor-intensive task, which must be repeated every time the detector changes. Recently, several groups have sought to sidestep this bottleneck by replacing components of the simulation pipeline with machine learning models that emulate the replaced components \cite{Chen:2021gdz, Carminati:2020kym, Sergeev:2021adf, The_ATLAS_Collaboration2024-yf}. A significant benefit of models that permit extremely fast sampling from the conditional densities is that they provide a straightforward way to fold 
theoretical predictions so that these can be compared directly with unfolded generator level observations. This is of particular interest to those who wish to use published statistical models and likelihoods \cite{Cranmer:2021urp}, which typically use unfolded observations.

Fast folding is of interest even with unfolded data. In measurements of QCD differential observables such as jet differential cross sections in nuclear and particle physics, the observables are unfolded to remove detector effects (see, for example, \cite{CMS:2021yzl, The_ATLAS_collaboration2021-kq}). However, the QCD predictions are made at the parton level. Therefore, it is necessary to map the predicted differential observables at the parton level to the level of the unfolded observables. In this context, the ratio $o(\bm{x}) / u(\bm{x})$ is referred to as the non-perturbative correction to the theoretical prediction.

In this paper, we demonstrate the effectiveness of implicit quantile networks (IQN) in modeling
$p(\bm{y} | \bm{x})$. Given a large sample of paired simulated jets---one at the particle \emph{generation} level (i.e., before the particles enter the particle detectors) and the other at the \emph{event reconstruction} level (i.e., jets recorded in the detectors)---we show that the \emph{jet response function} $p(\bm{y} | \bm{x})$ can be accurately modeled with an IQN. We shall refer to the sampling operation of an IQN for nuclear and particle physics observables as \emph{stochastic folding}.

The paper is organized as follows. In Sec.~\ref{sec:relatedwork} we briefly describe related work and the work that inspired the current paper. This is followed in Sec.~\ref{sec:IQN} by a description of the IQN model. Section~\ref{sec:app} describes the datasets used, the training of our models, and the results. We discuss these results in Sec.~\ref{sec.discussion} and give our conclusions in Sec.~\ref{sec:conclusion}.

\section{Related Work}
\label{sec:relatedwork}

The use of neural networks to approximate $p(\bm{y} | \bm{x})$ was first studied by White \cite{White_1992} and Taylor \cite{taylor_2000}, an approach that is of considerable interest in many fields including  high-energy physics; see, for example, \cite{Baldi:2016fzo,Cranmer:2019eaq,Brehmer:2020cvb, gouttes2021probabilistic} and the references therein. Generative adversarial networks (GAN) \cite{Blue:2021lxz, 10.21468/SciPostPhys.8.4.070, Erdmann_2019, Paganini_2018},  normalizing flows \cite{Albergo_2019,Gao_2020,Gao_2020a,Bieringer_2021, caloFlow}
and combinations of flows, GANs, and autoencoders \cite{Brehmer:2020vwc} for modeling conditional densities
have been  explored in a variety of applications. Recent work has also proposed methods for calibrating predictive distributions \cite{dey2022calibrated}. Deep neural networks have been trained using the average \emph{quantile loss} (see Sec.~\ref{sec:IQN}) to model detector response and to perform jet reconstruction \cite{osti_1617153, Sirunyan2020}. In the latter works, however, researchers focused on accurately modeling specific quantiles of interest. In contrast, we model the full quantile function using deep neural networks by extending the network architecture to include the quantile $\tau$ as one of the inputs---an approach first described by Ostrovski \textit{et\,al.~} \cite{Ostrovski2018AutoregressiveQN}. A preliminary version of this work was presented at the Machine Learning and the Physical Sciences (ML4PS) workshop co-located with the Neural Information Processing Systems conference in 2021 \cite{kronheim_iqn}.



\section{Implicit Quantile Networks}
\label{sec:IQN}



Given a one-dimensional conditional cumulative distribution function $\tau = F(y | \bm{x})$, the \emph{quantile function} is its inverse $y = F^{-1}(\tau | \bm{x})$. The quantile function thus maps a given cumulative probability (quantile) $\tau$ to the value $y$ of the random variable $Y$ such that $\textnormal{Pr}[Y \leq y] = \tau$. Given a training dataset comprising samples $y$ conditional on $\bm{x}$, the \emph{quantile regression} problem is to construct an estimator $f(\tau, \bm{x}; \bm{\theta})$ parameterized by $\bm{\theta}$ that approximates $F^{-1}$. This problem can be cast as an optimization problem \cite{koenker_bassett_1978} in which the quantile loss,



\begin{equation}
    \label{eq:loss}
    \mathcal{L}(f, y) = \left\{
    \begin{array}{ll}
        \tau (y-f(\tau, \mathbf{x}; \pmb{\theta})) &  y \geq f(\tau, \mathbf{x}; \pmb{\theta})\\
        (1 - \tau)(f(\tau, \mathbf{x}; \pmb{\theta})-y) &  y <  f(\tau, \mathbf{x}; \pmb{\theta})
    \end{array}
    \right. ,
\end{equation}
averaged over a set of training examples $(\bm{x}, y)$ is minimized. An \emph{implicit quantile network} is a deep neural network that operates as the estimator $f$.

This method has many applications including the stochastic folding of one set of particles to another. We illustrate the simplicity and efficacy of the approach by applying it to the stochastic folding of jets at the generation level (``gen-jets'') to jets at the level of event reconstruction (``reco-jets''). Each gen-jet is defined by the $4$-momentum $\bm{x} = (p_T, \eta, \phi, m)$, where $p_T, \eta, \phi, m$ are the jet transverse momentum, pseudo-rapidity, azimuthal angle, and mass, respectively. A reco-jet is analogously defined by $\bm{y} = (p_T', \eta', \phi', m')$. Following Ostrovski \emph{et al.}, we can write the 4-dimensional conditional density $p(\bm{y} | \bm{x})$ as
\begin{align}
    p(\bm{y} | \bm{x})  & =  
    p(p_T'|\bm{x} )\nonumber\\
    &\times p(\eta'|\bm{x}, p_T' )\nonumber\\
    &\times p(\phi'| \bm{x}, p_T', \eta' )\nonumber\\
    &\times p(m' |  \bm{x}, p_T', \eta', \phi' ) .
\label{eq:cd}
\end{align}
Each of the four densities in Eq.\,(\ref{eq:cd}) can be modeled with
an independent IQN, as shown in Sec.\,\ref{sec:dljs}. In the following, we refer to this model as \iqnfour.
However, it is also possible to model Eq.\,(\ref{eq:cd}) with a \emph{single} IQN by a judicious choice of inputs.
Every training example $(p_T, \eta, \phi, m)\rightarrow(p_T', \eta', \phi', m')$  is ``unrolled'' into the four training examples


\begin{align}
(p_T, \eta, \phi, m,1,0,0,0,0,0,0) &\rightarrow  p_T',\nonumber\\
(p_T, \eta, \phi, m,0,1,0,0,p_T',0,0) &\rightarrow  \eta',\nonumber\\
(p_T, \eta, \phi, m,0,0,1,0,p_T',\eta',0)&\rightarrow \phi',\nonumber\\
(p_T, \eta, \phi, m,0,0,0,1,p_T',\eta',\phi')&\rightarrow m' ,
\label{eq:examples}
\end{align}

where the left-hand sides of
Eq.\,(\ref{eq:examples}) are the possible inputs to the single IQN and
the one-hot encoding after $p_T, \eta, \phi, m$ in Eq.\,(\ref{eq:examples}) specifies which target is associated with the given unrolled example. We call this model \iqnone. In addition to the one-hot encoding $\bm{z}$ and the differing components $\bm{y}$ of the reconstruction-level 4-momentum from the training sample, the quantile $\tau$ is also an input to our model $f(\tau, \bm{x}, \bm{z}, \bm{y}; \bm{\theta})$. During training, the quantile $\tau$ is repeatedly and independently sampled from $U(0, 1)$ and an independent value of $\tau$ is associated with each unrolled example. 

We train a deep neural 
network on batches of unrolled examples from the training set with randomly sampled $\tau$ values. At inference time, the trained model is used  autoregressively: the unrolled examples, each with a randomly sampled quantile, are provided in the order shown in Eq.\,(\ref{eq:examples}) with the quantities $p^\prime_T, \eta^\prime, \phi^\prime$ now the values \emph{predicted} by the trained model, rather than the values from the training data.

Since the trained model approximates four quantile functions---one quantile function at a time depending on which one-hot encoding is used---the model is an \emph{implicit} approximation of the multi-dimensional conditional density $p(\bm{y} | \bm{x})$ in the following sense. The 1-dimensional conditional densities, from which $p(\bm{y} | \bm{x})$ is formed using Eq.\,(\ref{eq:cd}), can be computed from $p(y^{(n)} | \bm{x}, y^{(1)}, \cdots, y^{(n-1)}) = (\partial f_n / \partial \tau)^{-1}$, where $f_n$ is the model specified with the $n^\textnormal{th}$ one-hot encoding. 

One problem that can arise in quantile regression is \emph{quantile crossing}, where the approximation to the quantile function is not monotonic. Prior work has attempted to mitigate this problem by imposing constraints on the model architecture or by optimizing novel loss functions\,\cite{cannon_2018, moon2021}. Tagasovska and Lopez-Paz, however, observed that the problem becomes significantly less pronounced when the full quantile function is approximated\,\cite{NEURIPS2019_73c03186}. In this paper, we propose using a regularized average loss 
$$ \mathcal{L}' = \mathcal{L} + \lambda \cdot \mathbbm{1}[-f'] (f')^2$$
that further alleviates this problem, 
where $f' = \partial{f}/\partial{\tau}$ and $\lambda$ is a hyperparameter. By penalizing negative gradients of $f$ with respect to $\tau$, the regularization term favors solutions that are monotonically non-decreasing. This is equivalent to the condition $ (f^\prime)^{-1} = p(y | \bm{x}) > 0$.

\section{Stochastic Folding of Jets}
\label{sec:app}
\subsection{Dataset Generation}
As this study is a proof of concept, we
limit our investigations to
the fast folding of gen-jets to reco-jets. 
The \texttt{Pythia8} \cite{Bierlich:2022pfr} event generator (v8.307) is used to simulate proton-proton collisions at 13\,TeV at the Large Hadron Collider (LHC)\footnote{\url{https://home.cern/science/accelerators/large-hadron-collider}}, while the \texttt{Delphes} \cite{Selvaggi:2016ydq} detector simulator (v3.5.0) provides an approximate simulation of the interaction of jets with one of the particle detectors at the LHC, namely, the Compact Muon Solenoid (CMS) detector \cite{CMS:2007sch}. 
A particle clustering algorithm, called anti-$k_\textnormal{T}$~\cite{Cacciari_2008}, is used at both the particle generation and event reconstruction levels to cluster particles into jets with a radius parameter $R=0.5$ using \texttt{FastJet} \cite{fastjet}.


We simulated $10$~million hard QCD jets using the \texttt{Pythia8}-\texttt{Delphes} pipeline, of which $1$~million were set aside as test data.  Of the remaining $9$~million jets, we used $8$~million as training data, leaving the remaining $1$~million jets for model validation and hyperparameter tuning. These jets had a generator level $p_T$ cut of 20 GeV and a reconstruction level cut of 5.0 on the absolute pseudorapidity corresponding to the edge of the CMS Hadronic Forward detector. Both tracker and calorimeter information was used for the particles included in the reconstruction level jets. The complete code to reproduce the results in this paper, released with a GPL3 license, can be found at the linked GitHub repository\footnote{\url{https://github.com/alpha-davidson/Jet-IQNs}}.

\subsection{Data Preprocessing}
\label{sec:data-proc}

Prior to training, the
simulated $4$-momenta  
are transformed as follows 

\begin{align} \label{eq:preprocess}
\mathbb{T}(p_T) & = z(\log p_T), \nonumber\\
\mathbb{T}(\eta) & = z(\eta), \nonumber\\
\mathbb{T}(\phi) & = z(\phi), \nonumber\\
\mathbb{T}(m) & = z(\log (m + 2)),
\end{align}

where $\mathbb{T}(.)$ denotes the transformed quantity and
$z(.)$ the function that standardizes its argument, i.e., ensures that the transformed quantity has zero mean and unit variance. 
We take the logarithm of $p_T$ and $m$  prior to standardization to reduce the range of these quantities, which can sometimes vary by orders of magnitude. The replacement of $m$ with $m+2$ in Eq.~(\ref{eq:preprocess}) avoids potential numerical problems with logarithms of small jet masses. We also transform the quantile $\tau$ provided to the IQN as follows

\begin{equation}
\mathbb{T}(\tau) = 6\tau - 3 ,
\label{eq:quantile-rescale}
\end{equation}
so that the transformed $\tau$ is on roughly the same scale as the other transformed inputs, while taking care to use the original value of $\tau$ in the computation of the loss function in Eq.\,(\ref{eq:loss}). Finally, the IQN   target was chosen to be

\begin{equation}
z\left(\frac{y_n + 10}{x_n + 10}\right), \qquad n = 1,\cdots,4,
\label{eq:normalization}
\end{equation}
where $x_n$ is the $n^{th}$ component of $(\log p_T, \eta, \phi, \log(m+2))$ and $y_n$ the corresponding component of $(\log p_T', \eta', \phi', \log(m'+2))$. 
In our experiments, we found this target to be easier to model than $z(y_n)$. The constant $10$ in the numerator and denominator of Eq.~\ref{eq:normalization} ensures that this ratio is always well-defined and positive. These transformations are appropriately inverted before the calculation of any downstream quantities of interest such as the marginal densities.


\subsection{Model Architecture and Training}
\label{sec:dljs}

Each individual network in the \iqnfour~model is a dense, feed-forward neural network. These models are completely independent of each other and can be trained in any order, or in parallel. The values of key hyperparameters---shared by all the networks---are listed in Table~\ref{tab:tf_network}. 
Each IQN is trained using AMSGrad \cite{amsgrad}, an optimization algorithm in the stochastic gradient descent family. We implemented a learning schedule, decreasing the learning rate of AMSGrad by a factor of $10$ after every $100$ epochs of training. When performing this annealing, we found it beneficial to resume training the model from the configuration in which it achieved the lowest validation loss over the past $100$ epochs, rather than from its final configuration. In our experiments, we achieved maximal performance with three such learning rate decay steps (i.e., with $400$ epochs of overall training). This process took about 24 hours using an Intel 14700KF with no core restrictions and 32G of RAM available. Note that the instantaneous CPU usage is usually around 2 cores and GPU acceleration was not useful for these models given the small network and batch sizes. 


As noted in Section~\ref{sec:IQN}, an alternative approach (\iqnone) is to model the 4-dimensional conditional density with a single, autoregressive network. 
Aside from operating on a different input representation (the unrolled examples presented in Equation~\ref{eq:examples}), the \iqnone~model is identical to the individual IQNs comprising the \iqnfour~model in every other way, including the network architecture, the training procedure, and duration.

\begin{table}[ht]
\begin{center}
\begin{tabular}{ | l | c | }
\hline
{\bf Hyperparameter} & {\bf Value } \\ 
\hline
number of layers & $5$ \\
\hline
nodes per layer & $50$ \\
\hline
parameter initialization & Glorot \cite{glorot} \\
\hline
activation function: LeakyReLU \cite{leakyrelu}&$\alpha=0.3$ \\
\hline
batch size & $512$ \\
\hline
initial learning rate & $10^{-3}$ \\
\hline
gradient penalty ($\lambda$) & $100$ \\
\hline
total training epochs & $400$ \\
\hline
\hline
\end{tabular}

\end{center}
\caption{Hyperparameter values used to configure and train the networks comprising the \iqnfour~and \iqnone~models.}
\label{tab:tf_network}
\end{table}

\subsection{Results}
\label{sec:results}

We assess the effectiveness of the trained models by comparing the predicted marginal density of each component of the reco-jet 4-momentum ($p^\prime_T, \eta^\prime, \phi^\prime, m^\prime$) with the corresponding reco-jet marginal density from the test set, both integrated over the full gen-jet phase space or over a small subset of the latter.

\begin{figure}[htb]
\begin{subfigure}{.5\textwidth}
\includegraphics[width=\linewidth]{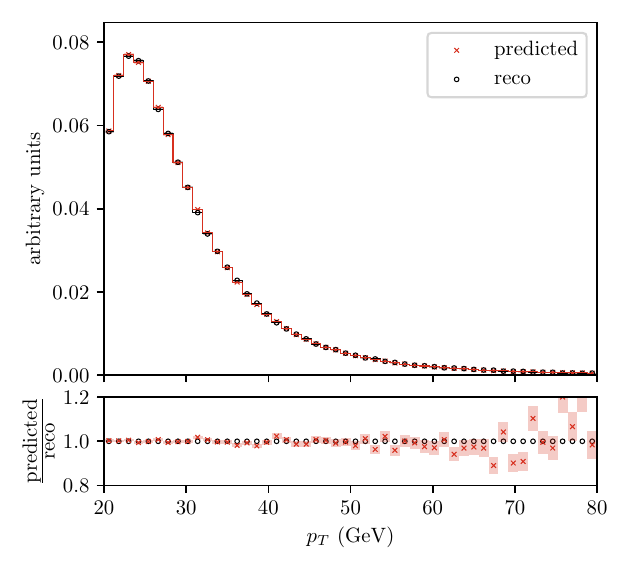}
\includegraphics[width=\linewidth]{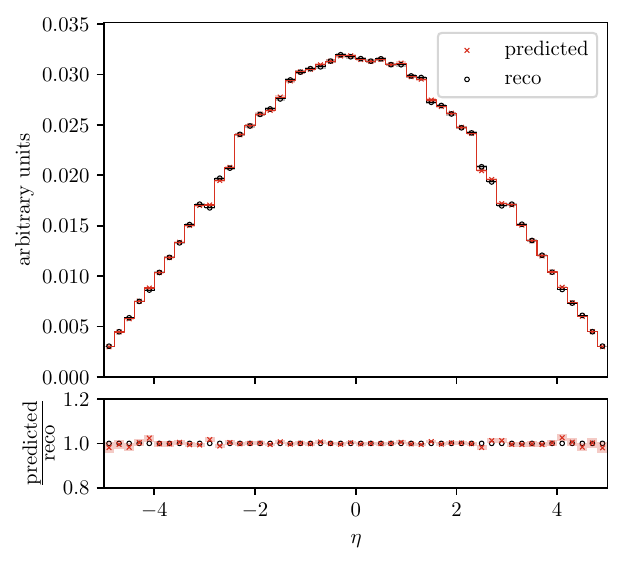}
\end{subfigure}%
\begin{subfigure}{.5\textwidth}
  \centering
  \includegraphics[width=\linewidth]{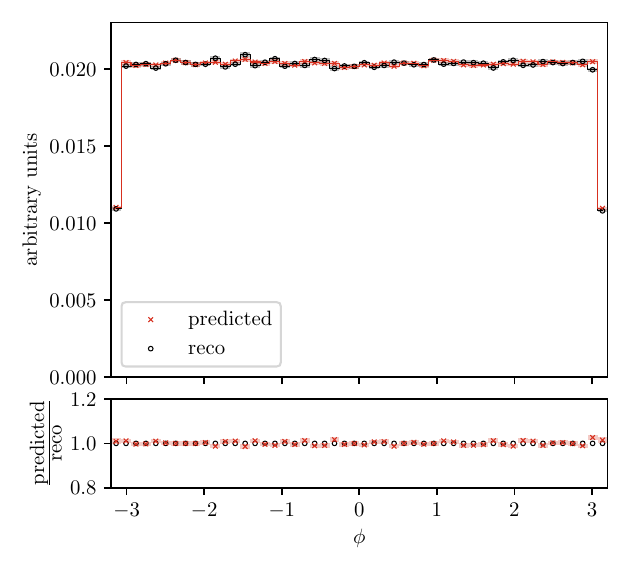}
  \includegraphics[width=\linewidth]{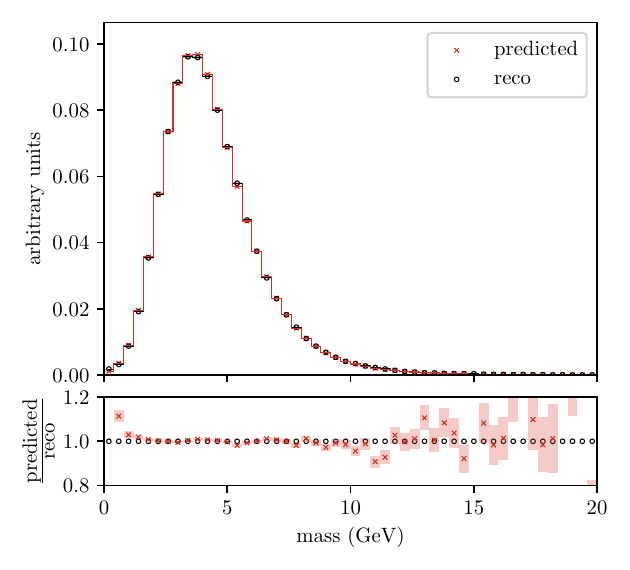}
  \end{subfigure}
\centering
\caption{Predicted marginal reco-jet spectra from the \iqnfour~model, one each for $p_T, \eta, \phi,$ and mass, superimposed on the corresponding original reco-jet distributions from the test set with a Poisson uncertainty. The ratios of the spectra are shown in the lower plots with the uncertainty propagated.
}
\label{fig:IQNx4}
\end{figure}

\begin{figure}[htb]
\begin{subfigure}{.5\textwidth}
  \centering
\includegraphics[width=\linewidth]{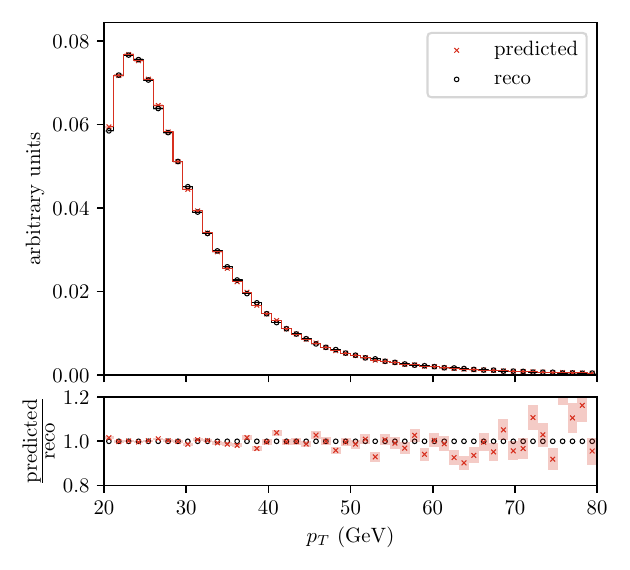}
\includegraphics[width=\linewidth]{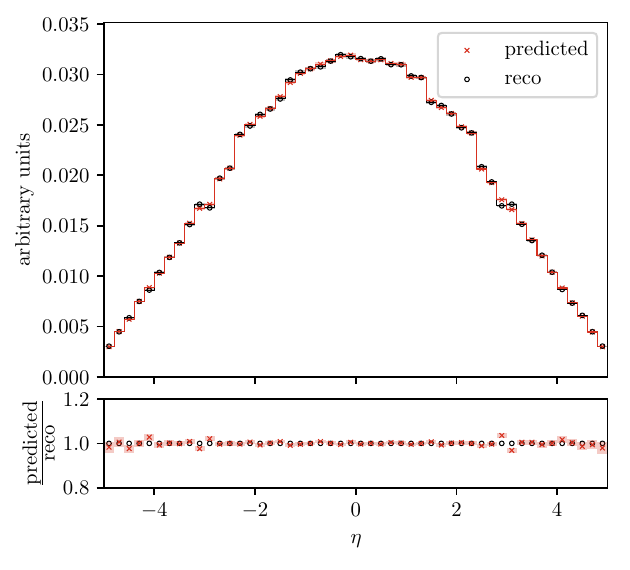}
\end{subfigure}%
\begin{subfigure}{.5\textwidth}
  \centering
  \includegraphics[width=\linewidth]{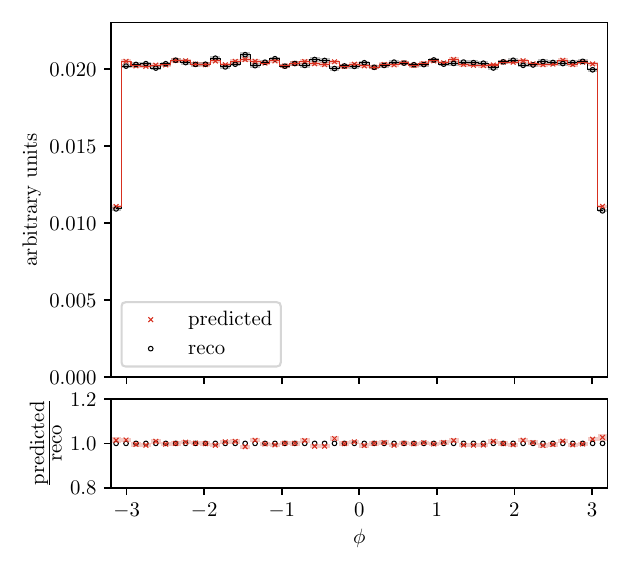}
  \includegraphics[width=\linewidth]{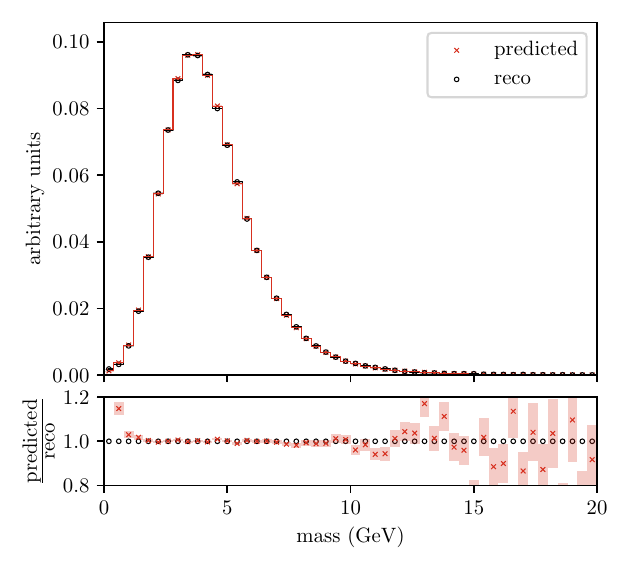}
  \end{subfigure}
\centering
\caption{Predicted marginal reco-jet spectra from the \iqnone~model for $p_T, \eta, \phi,$ and mass, superimposed on the corresponding original reco-jet distributions from the test set. The predicted spectra use one reco-jet prediction for each gen-jet in the test set with a Poisson uncertainty. The ratios of the spectra are shown in the lower plots with the uncertainty propagated. 
}
\label{fig:IQNx1}
\end{figure}

Using the trained \iqnfour~model, 1000 predictions for the reco-jet 4-momenta are generated for each of the $10^6$ gen-jets from the test set.
For each reco-jet 4-momentum component,
a single prediction is used per gen-jet to construct the reco-jet marginal densities. The reco-jet marginal densities are presented in Figure~\ref{fig:IQNx4} superimposed on the original reco-jet marginal densities computed from the test data. All uncertainties come from a normal approximation of the Poisson distribution to one standard deviation with a center value and variance equal to the number of counts in the bin. As described in Section~\ref{sec:IQN}, a single network, \iqnone, can be trained in an autogressive manner to accomplish the same task. For comparison, the same predictions are generated for this model in Figure~\ref{fig:IQNx1} as for the \iqnfour~model.



\begin{figure}[htb]
\begin{subfigure}{.5\textwidth}
  \centering
\includegraphics[width=\linewidth]{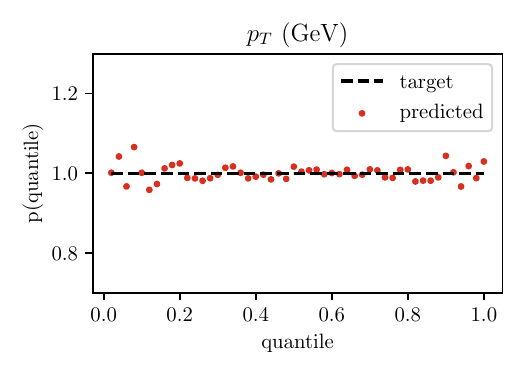}
\includegraphics[width=\linewidth]{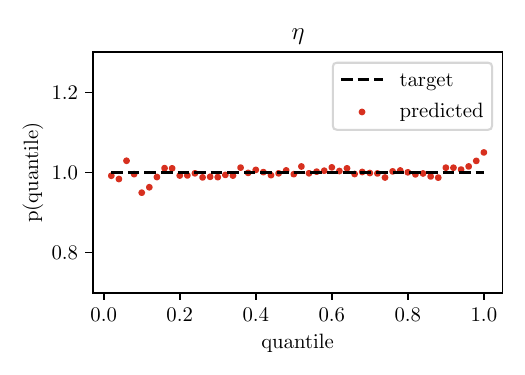}
\end{subfigure}%
\begin{subfigure}{.5\textwidth}
  \centering
  \includegraphics[width=\linewidth]{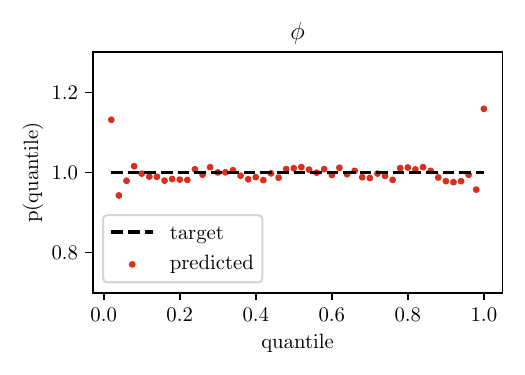}
  \includegraphics[width=\linewidth]{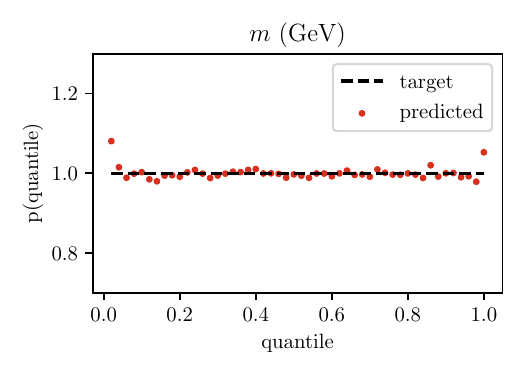}
  \end{subfigure}
\centering
\caption{The distribution of quantiles in the \iqnfour model, each computed from the reco-jet components in the test set using the trained model. 
}
\label{fig:all_closure}
\end{figure}

To further validate the \iqnfour, we perform the following closure test. The cumulative distribution function (cdf)  $\tau^\prime = F(y)$ for each component $y \in \{p^\prime_T, \eta^\prime, \phi^\prime, m^\prime\}$ of the reco-jet 4-momentum
is approximated using the 1000 predictions for each of the $10^6$ gen-jets in the test set.  The approximated cdfs are used to map each 4-momentum component $y$ in the test set to a quantile, $\tau^\prime$. If the IQN's modeling of the quantile function and, therefore, the conditional densities is accurate, then we should expect the distribution of the quantiles to follow $U(0, 1)$.  While there is some structure in the distributions in Fig.~\ref{fig:all_closure}, we see that the quantile distributions are indeed approximately uniform. 

As a further test we trained a boosted decision tree (BDT) model using XGBoost \cite{xgboost} to distinguish between the true reco-jet distribution and the simulated one. We created a binary classification task, where one category of examples comprised a gen-jet 4-vector concatenated with the original reco-jet 4-vector as generated by \texttt{Delphes}. The other category of examples were constructed by concatenating a gen-jet 4-vector with a reco-jet 4-vector as predicted by our model.
The BDT was trained with a maximum depth of 6 and an early stopping patience of 1000 and the binary cross entropy loss. The training set was 60\% of the main test set, while the validation and test sets were each 20\% of the main test set. The resulting BDT had an AUC of 0.53. Thus, it was able to find a small difference between the two datasets. As the binary cross entropy learns the mean of a distribution, this indicates that there is a slight discrepancy in the mean of the learned distributions. Given the agreement in the closure tests, we attribute these small discrepancies to be driven by the tails of the distributions. While small in number these can have significant impacts on the mean, thus impacting it more than the quantiles.

\begin{figure}[htb]
\begin{subfigure}{.5\textwidth}
  \centering
\includegraphics[width=\linewidth]{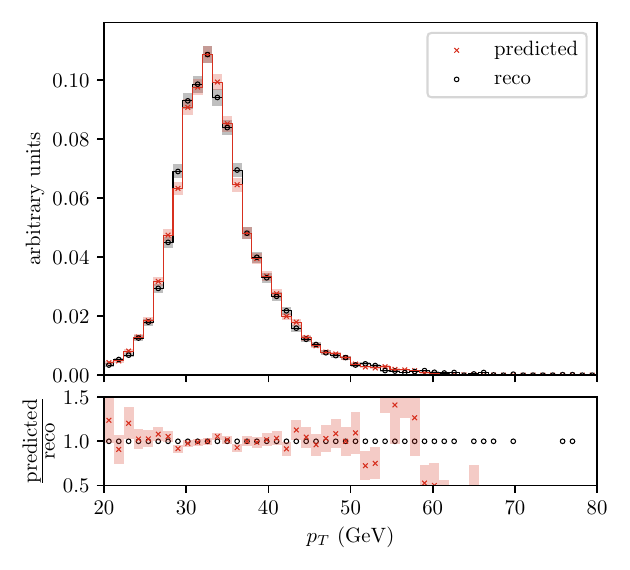}
\includegraphics[width=\linewidth]{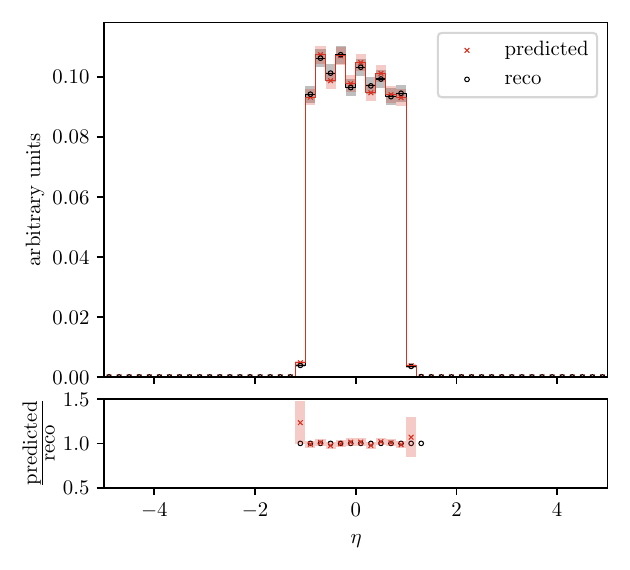}
\end{subfigure}%
\begin{subfigure}{.5\textwidth}
  \centering
  \includegraphics[width=\linewidth]{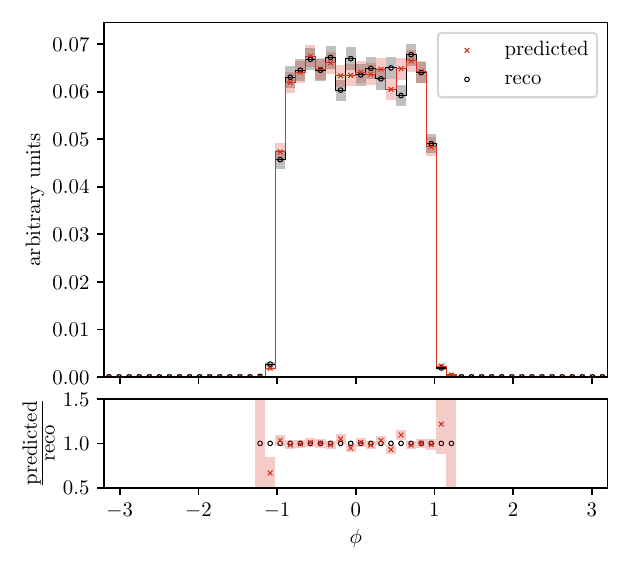}
  \includegraphics[width=\linewidth]{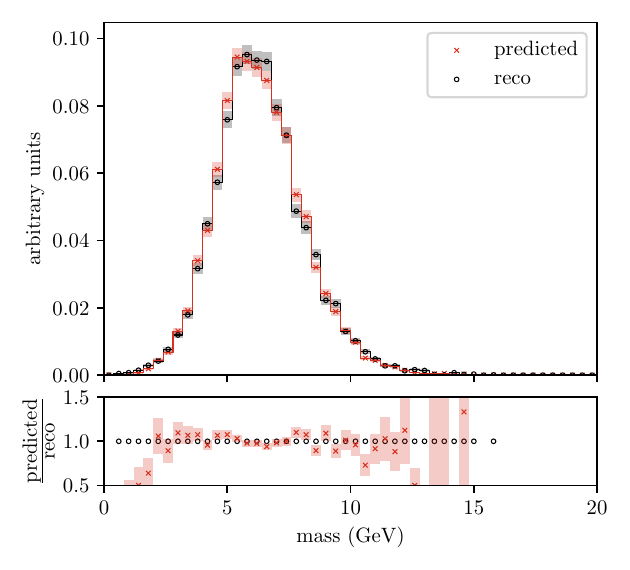}
  \end{subfigure}
\centering

\caption{The predicted marginal reco-jet spectra from the \iqnfour~model, 
conditioned on gen-jets in the phase-space bin $30\,\textnormal{GeV} < p_{T} < 35\,\textnormal{GeV}$, $|\eta| < 1$, $|\phi| < 1$, and $5\,\textnormal{GeV} < m < 10\,\textnormal{GeV}$, are compared to the reco-jet spectra of the test data. The predicted spectra displays the results of one prediction from the IQN for each gen-jet in the test set with a Poisson uncertainty. The ratios of the spectra are shown in the lower plots with the uncertainty propagated; $83\%, 92\%, 95\%,$ and $82\%$ of data points are displayed within chosen $y$-axis limits for $p_T, \eta, \phi,$ and mass respectively.}
\label{fig:cube_marg}
\end{figure}

The results presented above integrate over the entire gen-jet phase space. 
However, our claim is that IQNs are able to model conditional densities. We can investigate this claim by restricting
the gen-jet phase space to a small region.  The region defined by the criteria $30\,\textnormal{GeV} < p_{T} < 35\,\textnormal{GeV}$, $|\eta| < 1$, $|\phi| < 1$, and $5\,\textnormal{GeV} < m < 10\,\textnormal{GeV}$
contains approximately $1.2\times 10^4$ gen-jets with roughly the same 4-momenta. 
The associated distributions of reco-jet 4-momenta are shown in Figure~\ref{fig:cube_marg}. Again, we see excellent agreement between the predicted and true distributions. 

\begin{figure}[htb]

\includegraphics[width=\linewidth]{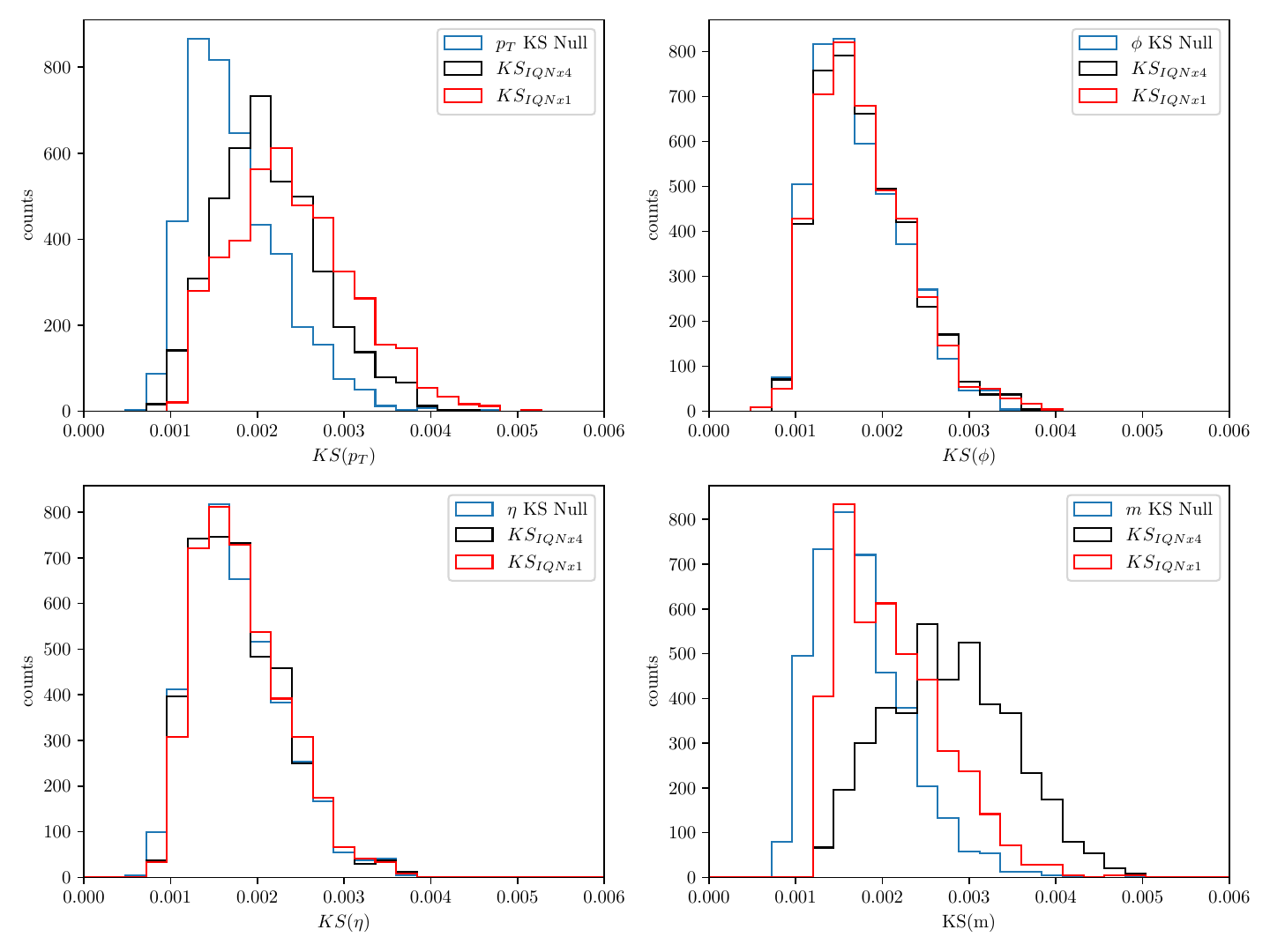}
\centering
\caption{Kolmogorov-Smirnow (KS) tests for each predicted variable computed between $1 \times 10^3$ tests of 1) two random sets of $5 \times 10^5$ samples each drawn from the test dataset (blue), 2) one random set of $5 \times 10^5$ samples from the test dataset and one random set of $5 \times 10^5$ samples from the \iqnfour~generated dataset (black), and 3) one random set of $5 \times 10^5$ samples from the test dataset and one random set of $5 \times 10^5$ samples from the \iqnone~generated dataset (red). }
\label{fig:KS_test}
\end{figure}

In order to further assess the effectiveness of our IQN methods, we compare the predicted reco-jet marginal densities to the original reco-jet marginal densities by performing two-sample tests using the Kolmogorov-Smirnov (KS) test statistic
\begin{equation}
    d_{\mathrm{KS}}(P, Q) \equiv \sup _{x \in \mathbb{R}}\left| F_{P}(x)-F_{Q}(x) \right|,
\end{equation}
where $F_P$ and $F_Q$ denote the cumulative distribution functions of each of the reco-jet variables of the samples $P$ and $Q$, respectively. The null distribution of the KS statistic is approximated by repeatedly splitting the test set into two sets $P_i$ and $Q_i$ using bootstrap resampling and computing the KS statistic for each pair of bootstrap samples $(P_i, Q_i)$. Similarly, the KS distribution for the \iqnfour~and \iqnone~models are computed in the same manner as the null distribution, but instead compares bootstrapped datasets from the null distribution and bootstrapped data from generated data from each model.
The results are shown in Figure~\ref{fig:KS_test} for both IQN models and compared with those of a baseline calculation using the simulated reco-jet distributions of the test set. The overlap in distributions of KS test results between the baseline distribution and our models indicate that our IQN models generate distributions similar to the baseline distributions. Recent work \cite{Das:2023ktd} proposes robust tests of generative models that will be interesting to apply to the IQN model for further analysis.

\section{Discussion}
\label{sec.discussion}
We have demonstrated the efficacy of IQNs for sampling from conditional densities. Moreover, the same model can be used to approximate the densities themselves. We have demonstrated that one can model the conditional probability distributions in the output space $\bm{y}$ by four independent deep neural networks. We also demonstrated that one can train a single network to accomplish the same task. The \iqnfour~and \iqnone~models yield excellent results over most of the jet phase space with small deviations where we expect to see them, namely, where there are comparatively fewer training data for $p_T$ and $m$, or where the modeling of detector shape effects in $\eta$ is difficult. It is likely, however, that these effects could be reduced by suitable adjustments to the training data, for example, by modeling the residuals $(\bm{\delta y} = \bm{y} - \bm{x})$  rather than the ratios. 

The choice of using a model in the style of \texttt{IQNx1}~ or \texttt{IQNx4}~ would depend on the specific use case. Based on the results obtained here the single model version is slightly more performant. This is likely due to information shared between the four different prediction modes, but this may not always be the case. Indeed, if independence between the different predictions is absolutely required, having separate network is a necessity as there is always the potential for small correlations in the single model method. Thus, the choice of model should be made based on the specific use case and its requirements.
 
 We also introduced a novel regularization term that mitigates quantile crossing without interfering with the convergence of the network training. For the particular tasks considered, we found that this term did not have a measurable impact. It is still in principle useful though, and may be of use in tasks with more complex distributions or with less training data. 

Conditional densities are ubiquitous in nuclear and high-energy physics. For example, they appear in statistical models $p(\bm{d} | \bm{\mu}, \bm{\nu})$, where $\bm{d}$ are observable data and $\bm{\mu}$ and $\bm{\nu}$ are parameters of interest and nuisance parameters, respectively. They also appear in response functions $r(\bm{y} | \bm{x}) $ that appear in multi-dimensional integrals of the form $o(\bm{y}) = \int r(\bm{y} | \bm{x}) \, u(\bm{x}) \, d\bm{x}$ that
map an unobserved spectrum $u(\bm{x})$ to an observed spectrum $o(\bm{y})$ of which the jet 4-momenta spectra are a typical example. Normalizing flows have been used for applications similar to the one addressed in our work, such as in \cite{caloFlow, Gao_2020a, Vaselli_2024, Bieringer_2021} with \cite{caloFlow} being the closest to our task. Preliminary work indicates their usefulness for our stochastic folding task as well. In these tests, more model choices and tuning are required than for the IQN approach, thus more targeted future work will be needed for a fair quantitative comparison. 

There is a renewed push in high-energy physics to publish full statistical models \cite{Cranmer:2021urp}. IQNs provide a simple and effective way to both encapsulate statistical models and to compute them quickly, as well as to model the numerous response functions that appear in the analysis of particle physics data at the Large Hadron Collider and other particle physics research facilities.

Finally, IQNs could be the basis of very fast simulators in which the hand-coding of conditional densities is replaced by appropriately trained networks. A high-fidelity simulator, such as the ones based on \texttt{GEANT4}, can be regarded as a tree of conditional densities from which one samples. The slower parts of a full \texttt{GEANT4}-based simulator could be replaced by fast emulators modeled using IQNs. Indeed, this is precisely the motivation for fast simulators like \texttt{Delphes}, which is itself outperformed in speed by our IQN-based approach by a factor of $\approx 500$.


\section{Conclusions}
\label{sec:conclusion}

In this work, we presented an application of IQNs, namely, the stochastic folding of jet observables. Our IQN architecture comprises a feed-forward, fully-connected neural network, which is straightforward to train, particularly when compared to other generative modeling techniques such as GANs that tend to be more unstable and require more careful tuning. Our approach uses one-hot encoding to select which quantity of the multi-dimensional conditional density is computed and is conditioned on the desired quantile. Consequently, by randomly sampling quantiles from $U(0, 1)$, samples from the multi-dimensional density can be readily generated. The trained IQNs approximate the marginal densities of the jet observables accurately across the jet phase space. Furthermore, we provided some evidence that this is also true for conditional densities. But confirming that this is true point-by-point over the jet phase space is a challenging task that is the focus of ongoing study. 

\begin{ack}

This work was supported in part by the U.S. Department of Energy (DOE) under Award No. DE-SC0010102 (HP) and the National Science Foundation (NSF) under Grant No. PHY-2012865 (MPK and RR) and Cooperative Agreement OAC-1836650 (BK). This work was performed in part at the Aspen Center for Physics, which is supported by National Science Foundation grant PHY-1607611. 

\end{ack}



\bibliographystyle{unsrt}
\bibliography{main}

\end{document}